\documentstyle[prl,aps,multicol,graphicx,epsfig,citesort]{revtex}

\begin{document}
\title{Heavy-Fermion Formation at the Metal-to-Insulator Transition
in Gd$_{1-x}$Sr$_x$TiO$_3$}
\author{M. Heinrich, H.-A. Krug von Nidda, V. Fritsch, and A. Loidl}
\address{Institut f\"{u}r Physik, Universit\"{a}t Augsburg, 86135 Augsburg, Germany}
\date{\today}
\maketitle

\begin{abstract}
The perovskite-like transition-metal oxide
Gd$_{1-x}$Sr$_x$TiO$_3$ is investigated by measurements of
resistivity, specific-heat, and electron paramagnetic resonance
(EPR). Approaching the metal-to-insulator transition from the
metallic regime ($x \geq 0.2$), the Sommerfeld coefficient
$\gamma$ of the specific heat becomes strongly enhanced and the
resistivity increases quadratically at low temperatures, which
both are fingerprints of strong electronic correlations. The
temperature dependence of the dynamic susceptibility, as
determined from the Gd$^{3+}$-EPR linewidth, signals the
importance of strong spin fluctuations, as
observed in heavy-fermion compounds. \\


\end{abstract}

\begin{multicols}{2}
The transition-metal oxides $R$TiO$_3$ (where $R$ denotes
Y$^{3+}$, La$^{3+}$, or some trivalent rare-earth ion) are known
as typical Mott-Hubbard insulators with Ti$^{3+}$ in a 3$d^1$
electronic configuration \cite{Imada}. Doping $R_{1-x}A_x$TiO$_3$
with divalent alkali-earth ions $A$ like Ca$^{2+}$ or Sr$^{2+}$
partially changes the Ti valence to Ti$^{4+}$ (3$d^0$) and induces
metallic behavior at a certain doping concentration $x$. The
metal-to-insulator transition of La$_{1-x}$Sr$_x$TiO$_3$ and
Y$_{1-x}$Ca$_x$TiO$_3$ has been investigated in detail
\cite{Tokura,Taguchi}. A critical increase of the Sommerfeld
coefficient $\gamma$ of the specific heat approaching the
metal-to-insulator transition has been observed in both compounds
with $\gamma \approx 25$\,mJ/(mole\,K$^2$) in
Y$_{0.6}$Ca$_{0.4}$TiO$_3$, indicating an enhancement of the
effective electron mass. Heavy-fermion formation has been observed
in a number of transition-metal oxides, with
Sr$_{1-x}$Ca$_x$RuO$_3$ \cite{Yoshimura} and LiV$_2$O$_4$
\cite{Kondo} being the most prominent examples. In the latter two
compounds the temperature dependence of the dynamic
susceptibility, as probed by the spin-lattice relaxation time in
NMR experiments, played a key role in identifying the
heavy-fermion behavior. In this letter we investigate the dynamic
susceptibility utilizing Gd EPR and provide experimental evidence
of heavy-fermion behavior in Gd$_{1-x}$Sr$_x$TiO$_3$.

Ceramic samples of Gd$_{1-x}$Sr$_x$TiO$_3$ have been calcinated
from TiO$_2$, SrCO$_3$, Gd$_2$O$_3$ and Ti$_2$O$_3$ powders of
high purity (better than 3N) and pressed into pellets. These
pellets have been annealed at 1473\,K under N$_2$ atmosphere for
15\,hours and finally arc-melted under argon atmosphere. For Sr
concentrations $x \leq 0.7$ X-ray powder diffraction revealed the
proper orthorhombically distorted GdFeO$_3$ perovskite structure
\cite{Reedyk}. For $x > 0.7$ the system attains the cubic
perovskite structure like pure SrTiO$_3$. The volume of the unit
cell remains nearly constant at about 235\,$\AA^3$ in the whole
concentration range. Susceptibility measurements were performed
with a commercial SQUID magnetometer (Quantum Design) in a
temperature range 4.2 $\leq T \leq$ 400\,K. Pure GdTiO$_3$ shows a
ferrimagnetic susceptibility, which follows a Curie-Weiss law
$\chi \propto (T+\Theta)^{-1}$ with $\Theta = 10$\,K at high
temperatures $T > 100$\,K, and exhibits an ordering temperature
$T_C \approx 28$\,K. This fits reasonably to the value of 34\,K
given in literature \cite{Greedan}, where the ordered state is
reported to show an antiparallel alignment of the
ferromagnetically ordered Ti$^{3+}$ spins with respect to the
Gd$^{3+}$ spins. With increasing $x$ the Curie-Weiss temperature
$\Theta$ shows approximately a linear decrease and vanishes for
$x \rightarrow 1$, whereas the magnetic order strongly becomes
suppressed. Already at $x > 0.1$ no magnetic order was observed
for $T \geq 1.7$\,K from our EPR measurements reported below. For
all samples, the paramagnetic susceptibility is dominated by the
contribution of the Gd$^{3+}$ spins ($S_{\rm{Gd}}$ = 7/2), which
is by a factor
$S_{\rm{Gd}}(S_{\rm{Gd}}+1)/S_{\rm{Ti}}(S_{\rm{Ti}}+1)$ = 21 times
stronger than the contribution of the Ti$^{3+}$ spins
($S_{\rm{Ti}}$ = 1/2). Comparing the theoretically expected and
experimentally observed susceptibility, we obtained that the
samples exhibit the correct composition $x$ within an error of
$\Delta x \approx 0.02$.

Electrical resistivity, using standard four-probe lock-in
technique, and specific heat, using the adiabatic method, were
measured in home-built $^4$He cryostats in a temperature range 1.5
$\leq T \leq$ 300\,K on bulk samples cut from the arc-melted
ingots. Fig.~1 shows the temperature dependence of the resistivity
for Sr concentrations $x \leq 0.5$ normalized to the value at $T
= 300$\,K. For pure GdTiO$_3$ the resistivity strongly increases
on decreasing temperature, as it is expected for a Mott insulator.
With increasing Sr concentration the resistivity strongly
decreases and a metal-to-insulator transition occurs for $x
\approx 0.2$. Within the concentration range 0.3 $\leq x \leq$ 0.5
the resistivity obeys roughly a quadratic temperature dependence,
which is demonstrated in the inset of Fig.~1. The observed
temperature dependence $\rho - \rho_0 \propto T^2$, instead of the
usual phonon contribution $\rho - \rho_0 \propto T^5$, is typical
for electron-electron scattering and only observable, if the
effective mass of the electrons is strongly enhanced due to
electronic correlations like in heavy-fermion systems \cite{Ott}.
At Sr concentrations $x > 0.6$ (not shown in Fig.~1) the system
becomes gradually insulating, approaching the band insulator
SrTiO$_3$ \cite{Cardona}.
\begin{figure}
\includegraphics[clip, width=8cm]{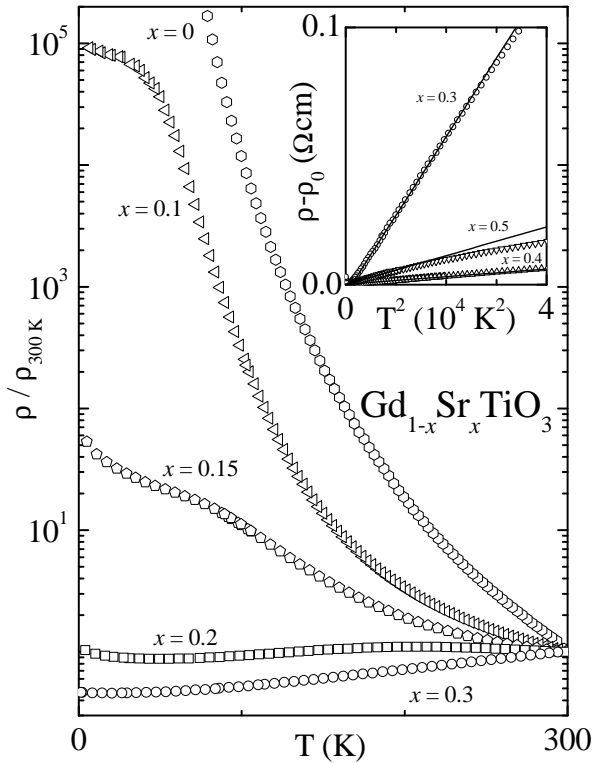}
\caption{Temperature dependence of the resistivity $\rho(T)$ for
Gd$_{1-x}$Sr$_x$TiO$_3$, 0$ \leq x \leq$ 0.3, normalized to the
value at $T = 300$\,K. Inset: Resistivity as a function of $T^2$
for 0.3$ \leq x \leq$ 0.5 after subtraction of the
zero-temperature value $\rho_0$. The solid lines indicate a
quadratic temperature dependence.} \label{fig1}
\end{figure}

Fig.~2 shows the temperature dependence of the specific heat $C$
for selected samples. Using the representation $C/T$ as a function
of $T^2$, the linear increase is due to the phonon contribution
$\beta$, following $C_{\rm{ph}} = \beta T^3$, and its
extrapolation to $T = 0$ yields the Sommerfeld coefficient
$\gamma$ due to the electronic contribution $C_{\rm{el}} = \gamma
T$. The increase of the data below 10\,K is due to the onset of
magnetic order of the Gd spins below 1\,K, as we proved for $x =
0.4$ in a $^3$He/$^4$He-dilution refrigerator. From the
extrapolation of the data above 10\,K down to $T = 0$ we
determined the Sommerfeld coefficient $\gamma$ for the metallic
regime, which is shown in the inset of Fig.~2. Approaching the
metal-to-insulator transition from the metallic side, the
Sommerfeld coefficient is strongly enhanced up to 50 times with
respect to simple metals at $x = 0.2$. This is another hint
concerning the enhancement of the effective electronic mass $m^*$
or - in the same sense - of the density of states $N(E_{\rm{F}})$
at the fermi energy, because $\gamma \propto m^* \propto
N(E_{F})$. Comparing these results to resistivity and specific
heat in the well investigated systems La$_{1-x}$Sr$_x$TiO$_3$ and
Y$_{1-x}$Ca$_x$TiO$_3$, we find a quite good agreement. An
enhancement of the electronic density of states at low
temperatures around the Mott-Hubbard transition is also predicted
by theoretical calculations \cite{Bulla}. Hence
Gd$_{1-x}$Sr$_x$TiO$_3$ is an appropriate system to study the
dynamic susceptibility close to a
\begin{figure}
\includegraphics[clip, width=8cm]{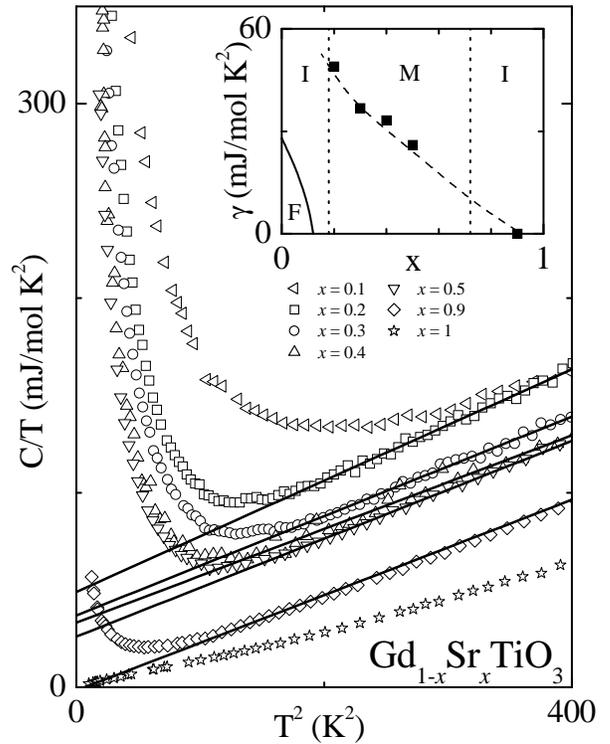}
\caption{Temperature dependence of the specific heat $C(T)$
plotted as $C(T)/T$ versus $T^2$ for Gd$_{1-x}$Sr$_x$TiO$_3$. The
solid lines indicate a temperature dependence $C(T) = \gamma T +
\beta T^3$. The pure compound ($x = $0) was omitted, as it is
dominated by magnetic order in this temperature regime. Inset:
Sommerfeld coefficient $\gamma$ as a function of the Sr
concentration $x$.} \label{fig2}
\end{figure}
Mott-Hubbard transition.

The EPR experiments have been performed at a Bruker ELEXSYS
E500-CW spectrometer at X-band frequencies (9.4\,GHz) equipped
with continuous gas-flow cryostats for He (Oxford Instruments) and
N$_2$ (Bruker) in the temperature range between 4.2 and 680\,K.
For temperatures down to 1.7\,K we used a cold-finger $^4$He-bath
cryostat. To avoid the influence of the skin effect, which
prevents the microwave to enter metallic samples, the arc-melted
ingots were powdered, filled in quartz tubes, and embedded in
paraffin or NaCl for temperatures below or above room temperature,
respectively.

The EPR spectra of all compounds consist of a single
exchange-narrowed resonance line of Lorentzian shape with a g
value $g \approx 2$. The evaluation of comparable Gd$^{3+}$
spectra has been described in detail in a previous publication
\cite{Krug}. Their intensity follows the Curie-Weiss behavior of
the static susceptibility measured with the SQUID magnetometer and
comparison with a standard sample showed that it is due to all
Gd$^{3+}$ spins. In the Mott-insulating regime $x \leq 0.15$ the
resonance linewidth $\Delta H$ strongly increases with decreasing
temperature below $T < 50$\,K, due to the onset of magnetic
order. In the minimum near 50\,K, it attains a value of about
1\,kOe. To higher temperatures $T > 50$\,K the linewidth
increases with a slightly negative curvature and approximately
2.5\,Oe/K. In the metallic regime $0.2 \leq x \leq 0.6$, the
linewidth $\Delta H$ exhibits a clearly different behavior, which
is shown in Fig.~3: Starting at low temperatures the linewidth
strongly increases, develops a prominent maximum and decreases
again. At temperatures above 150\,K the linewidth remains nearly
constant ($x \leq 0.4$) or shows a further slight decrease ($x
\geq 0.5$). Approaching the metal-to-insulator transition from
above ($x \geq 0.2$), the maximum increases and shifts to lower
temperatures.

In the following discussion, we confine ourselves to the metallic
regime: In metals the transversal spin-relaxation time $T_2$
equals the longitudinal or spin-lattice relaxation time $T_1$.
Hence the EPR linewidth $\Delta H \propto 1/T_{2}$ directly
measures the spin-lattice relaxation rate $1/T_1$, which is
usually found linear in temperature for simple metals, probing the
electronic density of states $N(E_{\rm{F}})$ due to the Korringa
law \cite{Barnes}:
\begin{equation}
\Delta H = b \cdot T, \hspace{0.2cm} {\rm where} \hspace{0.3cm} b
\propto N^2(E_{\rm{F}}). \label{Korringa}
\end{equation}
However, the temperature evolution of the linewidth observed in
Gd$_{1-x}$Sr$_x$TiO$_3$ differs completely from the linear
Korringa law \cite{Barnes}, but it reveals similar features as the
Gd ESR \cite{Krug,Elschner} and NMR spin-lattice relaxation rate
\cite{Asayama} in heavy-fermion compounds. Moreover comparable
NMR results have been reported form transition-metal oxides with
heavy-fermion formation like LiV$_2$O$_4$ \cite{Kondo} and
Ca$_{1-x}$Sr$_x$RuO$_3$ \cite{Yoshimura}. In all these systems a
second important contribution to the spin-lattice relaxation
arises from spin fluctuations of the electronic bath due to the
fluctuation-dissipation theorem, which for temperatures large
compared to the exciting frequency $k_B T \gg \hbar \omega$ reads
\begin{equation}
\frac{1}{T_1} \propto \frac{T}{\omega} \cdot {\rm{Im}}
\chi(\omega, T). \label{FDT}
\end{equation}
Here Im$\chi(\omega, T)$ denotes the imaginary part of the dynamic
susceptibility of the band states represented by the Ti $3d$
electrons in the present case. An appropriate expression for the
dynamic susceptibility, which takes the itinerant character of
the Ti electrons into account, is given by Ishigaki and Moriya
\cite{Ishigaki,Moriya}, who developed a theoretical description
of nuclear magnetic relaxation around the magnetic instabilities
in metals. Their results were also used to explain the NMR
relaxation rate in Ca$_{1-x}$Sr$_x$RuO$_3$ \cite{Yoshimura}.
Ishigaki's simulations of the nuclear spin-lattice relaxation
rate $1/T_1$ near a ferromagnetic instability - the Ti spins
order ferromagnetically (!) - look quite similar to our linewidth
data. Substituting the hyperfine coupling of the nuclear spin
with the electronic system by the super-exchange interaction
between Gd$^{3+}$ and Ti$^{3+}$ spins, we achieved an analogous
expression for the EPR linewidth. Moreover it was necessary, to
include the effect of the external magnetic field onto the
magnetic instability \cite{Moriya}. Finally we used the following
expression to fit our data:
\begin{equation}
\Delta H = A_{\rm{SE}} \cdot \frac{3 t y^2}{2(y^3 + h)} + \Delta
H_0. \label{Moriya}
\end{equation}
The reduced inverse 3$d$ susceptibility $y$ is given by an
implicit integral equation \cite{Ishigaki}, which has to be
iterated numerically. It is characterized by two parameters $y_0$
and $y_1$, respectively: $y_0$ is proportional to the inverse
static susceptibility at zero temperature and therefore vanishes
at the magnetic instability. $y_1 \equiv T_0 \cdot y_2$ is
proportional to the the energy width $T_0$ ($\sim 0.01-0.1$\,eV)
of the dynamical spin-fluctuation spectrum. Both parameters are
normalized to the width $T_{\rm{A}}$ ($\sim 1$\,eV) of the
distribution of the static susceptibility in momentum space. The
variable $t = T/T_0$ denotes the reduced temperature. The
parameter $h \propto H^2$ describes the influence of the applied
magnetic field $H$. The super-exchange coupling, which is included
in the prefactor $A_{\rm{SE}}$, is assumed to be temperature
independent. The residual linewidth $\Delta H_0$ is due to the
contribution of impurities and inhomogeneties in the sample. Hence
there are six parameters altogether to describe the temperature
dependence of the linewidth.

To obtain a reasonable description of our data, we tried to find a
parameter set, where the parameters $h$ - the resonance is found
at about $H = 3.4$\,kOe for all compounds - and $y_2 \propto
(1/T_{\rm{A}})^2$, which is related to the width of the conduction
band, attain the same values for all Sr concentrations. The
coupling constant $A_{\rm{SE}}$ and the residual linewidth $\Delta
H_0$ were allowed to change slightly. The most important effect
was expected to appear in the parameter $y_0$, which is directly
tuned by the Sr concentration $x$, and the energy width $T_0$ of
the spin-fluctuation spectrum. The results are drawn as solid
lines in Fig.~3 and the respective fit-parameters are listed in
Table~1. The fit nicely describes the experimental data at
temperatures $T \ll T_0$ and underlines the dominant influence of
the 3$d$-spin fluctuations for relaxation near the Mott-Hubbard
transition: Indeed the super-exchange coupling and residual
linewidth change only slightly within the metallic concentration
regime under consideration. Both parameters $y_0$ and $T_0$
increase nearly linearly with the Sr concentration $x$. The
critical concentration for the onset of magnetic order ($y_0 = 0$)
accompanied by the transition to the Mott insulator is found at $x
\lesssim 0.2$ in agreement with the resisitivity measurements.

The resonance linewidth at temperatures far below the maximum,
which is shown in the upper inset of Fig.~3, exhibits deviations
from the linear behavior predicted by the theory. The slope is too
small, and for $x \leq 0.4$ one observes even a slight upturn with
decreasing temperature instead of the expected linear decrease.
These deviations are probably due to the influence of the
crystal-electric field or the onset of magnetic order below $T =
1$\,K, which both induce a broadening of the resonance line. In
the case of a small crystal-electric field the linewidth at low
\begin{figure}
\includegraphics[clip, width=8cm]{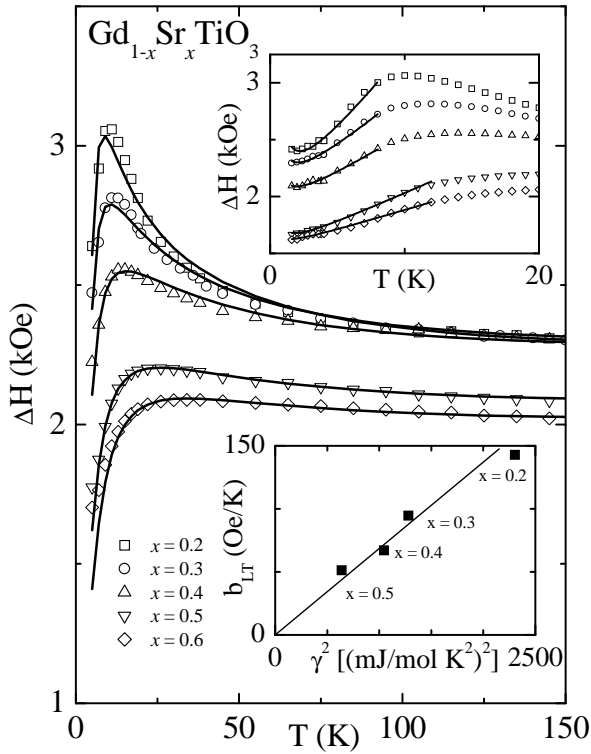}
\caption{Temperature dependence of the EPR linewidth $\Delta H(T)$
for Gd$_{1-x}$Sr$_x$TiO$_3$ within the metallic regime 0.2 $\leq x
\leq$ 0.6. The solid curves have been calculated from equation
(\ref{Moriya}). The fit parameters are given in table~1. Upper
inset: linewidth at low temperatures. The solid lines indicate a
temperature dependence following equation (\ref{CEF}). Lower
inset: Low-temperature Korringa slope $b_{\rm{LT}}$ as a function
of the squared Sommerfeld coefficient $\gamma^2$. The solid line
indicates the correlation of $b_{\rm{LT}}$ and $\gamma^2$.}
\label{fig3}
\end{figure}
temperatures follows (solid lines in the upper inset of Fig.~3)
\cite{Barnes}
\begin{equation}
\Delta H = b_{\rm{LT}} \cdot T + \frac{M_2}{b_{\rm{LT}} \cdot T} +
\Delta H_0, \label{CEF}
\end{equation}
where $M_2$ is proportional to the second moment of the
crystal-field splitting. It is interesting to compare the
low-temperature Korringa slope $b_{\rm{LT}}$ derived from equation
(\ref{CEF}) - having in mind the pure Korringa law
(\ref{Korringa}) - with the squared Sommerfeld coefficient
$\gamma^2$ of the specific heat, as is shown in the lower inset
of Fig.~3. The clear correlation of both parameters underlines the
enhancement of the electronic density of states $N(E_{\rm{F}})$ at
low temperatures in the metallic regime near the Mott-Hubbard
transition. It is visible by both macroscopic ($\gamma$) and
microscopic ($b_{\rm{LT}}$) probes.

In conclusion we have shown that the compound
Gd$_{1-x}$Sr$_{x}$TiO$_3$ exhibits a metal-to-insulator transition
near a Sr concentration $x = 0.2$: Electrical resistivity and
specific heat at the transition are in full agreement with the
well known systems La$_{1-x}$Sr$_x$TiO$_3$ and
Y$_{1-x}$Ca$_x$TiO$_3$. As Gd$^{3+}$ naturally belongs to the
compound, it is well suited for EPR investigations. The
temperature dependence of the Gd$^{3+}$ linewidth in the metallic
regime $0.2 \leq x \leq 0.6$ is determined by the Ti$^{3+}$-spin
fluctuations and is well described by the theory of Ishigaki and
Moriya for a metal near a ferromagnetic instability. The
low-temperature behavior of resistivity, specific heat and ESR
linewidth resembles to the properties of heavy-fermion systems,
but the enhancement of the effective electronic masses by a factor
50 is weaker than in f-derived heavy fermions, which are
dominated by Kondo-compensation effects and where one observes an
enhancement factor of 1000.

We are grateful to Dr. E. W. Scheidt, Experimentalphysik III,
University of Augsburg, for specific-heat measurements at
temperatures below 1\,K. This work was supported by the
Bundesministerium f\"ur Bildung und Forschung (BMBF) under the
contract number EKM 13N6917/0 and by the Deutsche
Forschungsgemeinschaft within Sonderforschungsbereich 484.

\begin{table}
\begin{center}
\begin{tabular}{ccccc}
$x$ & $y_0 (\cdot 10^{-4})$ & $T_0$[K] & $A_{\rm SE}$[kOe] &
$\Delta H_0$[kOe]
\\ \hline 0.2 & 0.1 & 570 & 0.47 & 0.00 \\
0.3 & 5 & 590 & 0.48 & 0.02 \\
0.4 & 11 & 610 & 0.46 & 0.15 \\
0.5 & 19 & 625 & 0.43 & 0.13 \\
0.6 & 24 & 640 & 0.43 & 0.10 \\
\end{tabular}
\end{center}
\caption{Parameters of the fit to the temperature dependence of
the EPR linewidth in Gd$_{1-x}$Sr$_x$TiO$_3$ following equation
(\ref{Moriya}) with constant values $y_2 = $0.0055\,K$^{-1}$ and
$h = 1.9 \cdot 10^{-9}$.}
\end{table}

\end{multicols}
\end{document}